\documentclass[aps,prd,onecolumn,nofootinbib,preprint,superscriptaddress]{revtex4-1}
\usepackage{hyperref}
\usepackage{ulem}
\usepackage{graphicx}				
\usepackage{amssymb}
\usepackage{amsfonts,amsmath,amssymb,graphics,epsfig}
\usepackage{comment}
\usepackage{footnote}
\usepackage{graphicx}
\usepackage{dcolumn}
\usepackage{bm}

\usepackage{geometry}                		
\usepackage{graphicx}				
\usepackage{amssymb}
\usepackage[T1]{fontenc} 
\usepackage{graphicx}
\usepackage{graphicx}
\usepackage{subcaption}
\usepackage{caption}
\usepackage{mwe}
\usepackage{amsmath}
\usepackage{slashed}
\usepackage{young}
\usepackage{amssymb}
\usepackage[usenames,dvipsnames]{color}
\usepackage{amsmath}
\usepackage{slashed}
\usepackage{young}
\usepackage{amsfonts}
\usepackage{hyperref}
\usepackage{amssymb}
\usepackage{footnote}
\usepackage[usenames,dvipsnames]{color}
\usepackage{natbib}

\newcommand{\dx}[1]{\text{d}#1}
\newcommand{\dd}{\text{d}}

\begin{document}
\title{Finite-Size Effects On The Self-Force}
\author{Yi-Zen Chu}
 \email{yizen.chu@gmail.com}
\affiliation{%
	Department of Physics, National Central University, Chungli 32001, Taiwan, \\
	Center for High Energy and High Field Physics (CHiP), National Central University, Chungli 32001, Taiwan
}%
\author{Klaountia Pasmatsiou}%
 \email{kxp265@case.edu}
\author{Glenn D. Starkman}%
 \email{glenn.starkman@case.edu}
\affiliation{%
Physics Department/CERCA/ISO \\ 
Case Western Reserve University Cleveland, Ohio 44106-7079, USA}%

\date{\today}

\begin{abstract}
Electromagnetic and linear gravitational radiation do not solely propagate on the null cone in $3+1$ dimensions in curved spacetimes, contrary to their well-known behavior in flat spacetime. Their additional propagation inside the null cone is known as the tail effect. A compact body will produce a signal whose tail will interact with its future worldline, thus producing a tail-induced self-force. We present new results for the tail-induced scalar, electromagnetic and gravitational self-force for a test mass in orbit around a central mass,  including effects from the internal structure of that body.
\end{abstract}

\maketitle

\newpage

\section{Introduction}

With the recent direct detection by LIGO \cite{Abbott_2017,:2016aa} of gravitational waves (GWs) from inspiraling binaries, there is now increasing interest in computing physical effects beyond the leading-order Keplerian dynamics of idealized point masses orbiting one another. 
Within the gravitational-waves literature, the so-called self-force problem has been an outstanding issue for several decades, and is usually studied in the case of a small body orbiting a larger central mass. 
The `extreme' versions of such systems are dubbed ``Extreme-Mass-Ratio-Inspiral" systems (EMRIs). 
These typically involve order solar-mass compact objects orbiting the order $10^6-10^9$ solar-mass supermassive black holes (SMBHs) that apparently reside within most, if not all, galactic centers. 
By the 2030's, the space-based gravitational-wave detector LISA \cite{Galley:2008ih,Gair_2017} will be launched to detect GWs of much lower frequencies than those to which the current LIGO and Virgo detectors are sensitive. 
Among LISA's primary targets are EMRIs. 
Hence, there is a need for a practical scheme to compute EMRI waveforms, with the associated self-force effects properly incorporated -- 
see, for instance, the reviews \cite{Barack:2018yvs} and \cite{Poisson:2011nh}.

In curved spacetimes, massless waves such as those of electromagnetism and gravitation no longer travel strictly on the null cone -- they also ``scatter'' off the background geometry and develop ``tails,'' propagating inside the light cone. At least within the de Donder gauge, this is responsible for the dominant gravitational self-force experienced by a compact body orbiting a SMBH: in the strong-gravity region of the latter, the former interacts with the tail portion of the signal it generated in the past. This, in turn, affects the orbital evolution of the system and the gravitational-wave signal that will be detected.

Abraham and Lorentz \cite{jackson} first calculated the recoil force on an accelerating charge caused by the emission of electromagnetic radiation. 
Dirac \cite{1938} derived the relativistic generalization of the Abraham-Lorentz force. 
Building on work by Hadamard \cite{hadamard}, it was DeWitt and Brehme \cite{DEWITT1960220} (followed shortly by Hobbs \cite{Hobbs}) who first pointed out the novel contribution to the electromagnetic self-force from the tail effect in curved spacetimes. 
(This has no analog in 4D Minkowski, where electromagnetic and linear gravitational waves propagate strictly on the light cone.) 
Mino, Sasaki and Tanaka \cite{Mino:1996nk} derived the linear gravitational self force for a point particle moving in an arbitrary background, where the metric satisfies Einstein's equations in vacuum.\footnote{For a treatment in non-vacuum spacetimes see \cite{Zimmerman:2014uja} and \cite{Zimmerman:2015rga}.} 
Quinn and Wald \cite{Quinn:1996am} derived similar results using an ``axiomatic approach." 
Nowadays, this linear tail-induced self-force vacuum-background equation carries the acronym MiSaTaQuWa.

In this work, we focus on the tail-induced self-force; not in the EMRI system, but where the central body is replaced by a significantly smaller mass so that the spacetime is weakly curved everywhere. 
This setup is more closely related to the post-Newtonian/Minkowskian description of the comparable-mass binary systems whose GWs are currently detected by LIGO (see for instance \cite{Poisson:1997ha}). 
In our setup, and at first order in perturbation theory, the tail arises from the null signal generated by the orbiting compact body scattering off the central mass and its Newtonian potential, before returning to exert a self-force on the orbiting body at a later time. 
In particular, we will improve upon the methods in DeWitt and DeWitt \cite{DeWitt:1964aa} as well as Pfenning and Poisson \cite{Pfenning:2000zf}, and show how to capture self-force effects that are sensitive to the internal structure of the central body. 
Our method allows the evaluation of these effects perturbatively in the ratio of the size of the central object to its separation from the point mass. 
We demonstrate this for a specific choice of the radial density profile. Comparison with the extant literature reveals finite-size self-force terms that had previously been overlooked. This should not be confused with the usual discussion on the quadrupole effects induced by tidal forces in the post Newtonian literature \cite{Blanchet:2013haa}. This is known to arise at 5 PN but here we are discussing the effect of the central body's intrinsic quadrupole moment on the orbiter's self-force. In this work, we show that for the scalar and electromagnetic case this enters at a lower PN order and comment on the potential appearance of finite size effects at 2 PN for the gravitational case.

Extended bodies have previously been considered in the literature. 
Isoyama and Poisson \cite{Isoyama:2012in} considered the self-force acting on a (scalar or electric) charge held in place outside a massive body. Harte considered extended sources moving in an arbitrary background spacetime \cite{Harte:2014wya} while
Harte, Taylor and Flanagan \cite{Harte:2018iim} considered the electromagnetic problem non-perturbatively. The body of work by Harte and collaborators lays out a formalism to compute scalar, electromagnetic and gravitational self-forces in various dimensions.  However, they did not appear to carry out any concrete computations of the finite size dependence of these self-forces within the context of binary systems.
Finally, Pfenning and Poisson \cite{Pfenning:2000zf} studied the gravitational self-force problem for point-like objects in the weak-field limit. 
They nominally considered an extended central object, but did not carry the relevant effects into the ultimate calculation of the self-force.

This manuscript is organized as follows: in section \ref{sectionfirst}, we review the generic equations of the self-force, while in \ref{sectionsecond}, \ref{sectionthird} and \ref{sectionfourth} we show the calculation involving the finite-size effects. 
After discussing our results in section \ref{Conclusions}, we provide the reader with more technical details in the Appendices.
 
\section{Self-force equations} \label{sectionfirst}

\noindent{\bf Scalar} \quad We consider a massive spin-zero test particle with trajectory $z^\mu(\tau)$ (where $\tau$ is the particle's proper time), moving in a background spacetime characterized by a metric $g_{\alpha\beta}$, the associated Ricci tensor $R_{\alpha\beta}$, and scalar $R$. 
The particle is coupled to a classical scalar field and is subject to an external force $f_\text{ext}^\alpha$.

The modified geodesic equation for this particle \cite{Quinn:2000wa} is\footnote{Throughout the manuscript we set $c=G_N=1$, unless specified otherwise.} 
\begin{equation}\label{scalarf}
m u_{\phantom{\alpha}; \beta}^\alpha u^{\beta} = f_\text{ext} ^\alpha + \frac{1}{3} \frac{q^2}{m} (\delta^\alpha_{\phantom{\alpha}\beta} + u^\alpha  u_\beta) \dot{f}_\text{ext}^\beta + \frac{1}{6} q^2 (R^{\alpha}_{\phantom{\alpha}\beta} + u^\alpha R_{\beta \gamma} u^\beta u^\gamma) +  f_\text{scalar} ^\alpha\,.
\end{equation}
This includes a ``self-force'' contribution 
\begin{equation}\label{eqn:fscalar}
f_\text{scalar} ^\alpha = q^2 (g^{\alpha \beta } + u^\alpha u^\beta) \int_{-\infty} ^{\tau^{-}} G_{,\beta}\left(z(\tau),z(\tau')\right) \dd\tau',
\end{equation}
where $u^\alpha(\tau) \equiv \dd z^\mu/\dd \tau$ represents the 4-velocity of the particle. 
The $\,_{,\beta}$ denotes partial derivative with respect to $z^\beta(\tau)$;
an overdot is the derivative  with respect to proper time; while $q$ is the scalar charge of the scalar.
$G$ is the scalar Green's function, obeying
\begin{align}
\label{GreensFunction_Scalar_Eqn}
(\Box- \xi R) G(x,x')  = -4\pi \frac{\delta^{(4)}(x-x')}{\sqrt[4]{g(x)g(x')}} .
\end{align}
As is conventional, $\xi$ is the non-minimal coupling of the scalar field to the Ricci scalar $R$.

Pioneering work by Hadamard \cite{hadamard} informs us that the retarded solution to the Green's function equation -- in a region of curved spacetime where the observer at $x$ can be linked via a unique geodesic to the spacetime point source at $x'$ -- is comprised of a term (proportional to Dirac's delta function) that propagates signals strictly on the light cone and another (proportional to a step function) that transmits signals within the null cone.
\begin{align}
\label{GreensFunction_Scalar_Soln}
G(x,x') =\Theta(t-t') \left( \sqrt{\Delta(x,x')} \delta(\sigma) + \Theta(-\sigma) V(x,x') \right)\,,
\end{align}

where,

\begin{align}
\label{VV determinant}
\Delta(x,x') =-\frac{ \det[\partial_{\mu '} \partial_{\nu '} \sigma_{x,x'}]}{|g g'|^{1/2}}\,,
\end{align}

is the Van Vleck determinant and V(x,x') obeys the homogeneous wave equation with the appropriate boundary conditions on the light cone. Here, the $\sigma$ is Synge's world function, half of the square of the geodesic distance between $x$ and $x'$, so that $\sigma=0$ is null while $\sigma < 0$ is timelike. 
It is the presence of the tail $V$ in \eqref{GreensFunction_Scalar_Soln} that gives rise to  \eqref{eqn:fscalar}. 
The integral in \eqref{eqn:fscalar} extends over the entire past history of the particle until ``almost'' the present time $\tau^{-}$. 
The $-$ indicates that the integral is only over the ``tail'' portion of the Green's function, and does not include the light-cone piece. 
We shall witness below, the same tail phenomenon is responsible for the analogous history-dependent contributions to the electromagnetic and linear-gravitational self-forces in curved spacetimes.

When the particle is moving in vacuum, $ R= 0$; and, if there are no external forces, $ f_\text{ext} = 0$. 
The remaining piece is the tail integral. 
Quinn derived (\ref{scalarf}) using an extended body coupled to a scalar field in the limit of small spatial extent\footnote{Here we are referring to the spatial extent of the orbiting body, not the central mass.}. 
An extensive discussion and delicate issues with this limit can be found in \cite{Quinn:1996am}.

\noindent{\bf Electromagnetism} \quad The electromagnetic force felt by a point charge is given by \cite{Quinn:1996am} 
\begin{equation}\label{EMf}
m u^\alpha_{\phantom{\alpha}; \beta} u^{\beta}= f_\text{ext} ^\alpha + \frac{2}{3} \frac{e^2}{m} (\delta^\alpha_{\phantom{\alpha}\beta} + u^\alpha  u_\beta) \dot{f}_\text{ext}^\beta + \frac{1}{3} e^2 (R^\alpha_{\phantom{\alpha}\beta} u^\beta + u^\alpha R_{\beta \gamma} u^\beta u^\gamma) +  f_\text{em} ^\alpha ,
\end{equation}
where the history-dependent self-force reads
\begin{equation} \label{eqn:fem}
f_\text{em} ^\alpha = -e^2  \int_{-\infty} ^{\tau^{-}} (G^\alpha _{\phantom{\alpha}\gamma' ; \beta} - G _{\beta \gamma'} ^{\phantom{\beta\gamma'};\alpha}) u^\beta  u^{\gamma'}\dd\tau'.
\end{equation}
Here and below, the primed index indicates the proper velocity is evaluated at the integration time, namely $u^{\gamma'} \equiv \dd z^\mu/\dd \tau'$. The Lorenz-gauge \footnote{At the level of the vector potential $A_\mu$, we mean $\nabla_\mu A^{\mu}=0$.} electromagnetic Green's function itself obeys
\begin{align}\label{waveeq.em}
\Box G_{\mu\nu'}(x,x') - R_\mu^{\phantom{\mu}\sigma} G_{\sigma\nu'}(x,x') 
= -4\pi g_{\mu\nu'}(x,x') \frac{\delta^{(4)}(x-x')}{\sqrt[4]{g(x)g(x')}} ,
\end{align}
where $g_{\mu\nu'}$ is the parallel propagator. Comparing (\ref{scalarf}) to (\ref{EMf}), we observe that the scalar case is technically simpler than its electromagnetic counterpart. Nevertheless their fundamental derivation follows the same rules.

\noindent{\bf Gravitation} \quad Finally, up to quadratic order in the point mass $m$, the gravitational force felt by a point mass is given by \cite{Mino:1996nk,Pfenning:2000zf,Quinn:1996am}
\begin{equation}\label{gravf}
m {u^\alpha}_{; \beta} u^{\beta}= f_\text{ext} ^\alpha - \frac{11}{3} m (\delta^\alpha_{\phantom{\alpha}\beta} + u^\alpha  u_\beta) \dot{f}_\text{ext}^\beta +  f_\text{grav} ^\alpha,
\end{equation}
where the tail-induced self-force is
\begin{equation} \label{eqn:fgrav}
 f_\text{grav} ^\alpha = -2 m^2  \int_{-\infty} ^{\tau^{-}} (2 G^\alpha _{\phantom{\alpha} \beta \mu ' \nu '  ;\gamma} - {G _{\beta \gamma \mu ' \nu '}} ^{;\alpha} + u^\alpha   {G _{\beta \gamma \mu ' \nu ' ; \delta}} u^\delta) u^\beta  u^{\gamma } u^{\mu ' } u^{\nu ' } \dd\tau'.
\end{equation}
The gravitational Green's function is related to the trace-reversed Green's function $\check{G}(x,x')_{\mu\nu \alpha'\beta'}$,  via
\begin{align}
\check{G}_{\mu\nu \alpha'\beta'}(x,x') 
&= \check{P}_{\mu\nu}^{\phantom{\mu\nu}\sigma\rho}(x) \check{P}_{\alpha'\beta'}^{\phantom{\alpha'\beta'}\lambda ' \kappa '}(x') G_{\sigma\rho \lambda'\kappa'}(x,x'), \\
\check{P}_{\mu\nu}^{\phantom{\mu\nu}\alpha\beta}(x) &\equiv \frac{1}{2} \left( \delta_\mu^\alpha \delta_\nu^\beta + \delta_\mu^\beta \delta_\nu^\alpha - g^{\alpha\beta} g_{\mu\nu} \right)\,.
\end{align}
 $\check{G}$ obeys the following vacuum (i.e., $R_{\mu\nu}=0$) equation in the de-Donder gauge \footnote{At the level of the trace reversed metric perturbation $\bar{\gamma}_{\mu \nu} $, we mean $\nabla^\mu \bar{\gamma}_{\mu \nu} =0$.} :
\begin{align}\label{waveeq.grav}
\Box \check{G}_{\mu\nu \alpha'\beta'}(x,x') + 2 R_{\mu \phantom{\alpha} \nu \phantom{\beta}}^{\phantom{\mu} \alpha \phantom{\mu} \beta} \check{G}_{\mu\nu \alpha'\beta'}(x,x') = -2\pi \left( g_{\mu \alpha'} g_{\nu\beta'} + g_{\mu\beta'} g_{\nu\alpha'} \right) \frac{\delta^{(4)}(x-x')}{\sqrt[4]{g(x)g(x')}} .
\end{align}
The equations of motion in all of the 3 cases (eqs. \eqref{scalarf}, \eqref{EMf}, and \eqref{gravf}) are integro-differential equations that require us to know the entire past history of the particle.

\section{The two-point functions}\label{sectionsecond}

Throughout the paper, we set  $f_\text{ext}=\dot{f}_\text{ext}=0$, so that we are left only with the integral over the entire history of the particle. We solve the wave equations (\ref{GreensFunction_Scalar_Soln}), (\ref{waveeq.em}), and (\ref{waveeq.grav}), perturbatively to first order in the Newtonian potential, in the weak-field limit of the Schwarzschild metric,
\begin{equation}
\label{metric1}
\dx{s}^2= - (1-2\Phi) \dx{t}^2 + (1+2\Phi) \delta_{ij} \dx{x}^i \dx{x}^j, 
\end{equation}
to obtain \cite{Pfenning:2000zf, Chu:2011ip} \footnote{Our $G^{(1)}$ is equivalent to $\dot{G}$ in (\cite{Pfenning:2000zf}).},
\begin{equation}
G(x,x') = G^{\rm flat}(x,x') + G^{(1)}(x,x') +  {\cal{O}}\left(\Phi^2 \right), 
\label{Gperturbscalar}
\end{equation}
\begin{equation}
G^\alpha_{\ \beta'}(x,x') = G^{\rm flat}(x,x') \delta^\alpha_{\ \beta'}
+ {G^{(1)}}^\alpha_{\ \beta'}(x,x') +   {\cal{O}}\left(\Phi^2 \right),
\label{3.17}
\end{equation}
\begin{equation} 
G^{\alpha\beta}_{\ \ \gamma'\delta'}(x,x') = \Bigl( 
\delta^{(\alpha}_{\ \gamma'} \delta^{\beta)}_{\ \delta'} 
- \frac{1}{2} \eta^{\alpha\beta} \eta_{\gamma'\delta'} \Bigr)  
G^{\rm flat}(x,x') 
+ {G^{(1)}}^{\alpha\beta}_{\ \ \gamma'\delta'}(x,x') +   {\cal{O}}\left(\Phi^2 \right),
\label{3.29}
\end{equation}
where the flat retarded Green's function is given by,
\begin{equation}\label{delta}
G^\text{flat} (x,x') =\frac{\delta(t-t' - |\vec{x}-\vec{x}'|)}{|\vec{x}-\vec{x}'|},
\end{equation}
and satisfies the wave equation with a delta-function source,
\begin{equation}
\partial^2 G^\text{flat} (x,x') =\delta^{(4)}(x-x').
\end{equation}
According to eq. (3.14) of \cite{Pfenning:2000zf},
\begin{equation}\label{Gscalarscalar}
G^{(1)}(x,x')  = -2 \partial_{t t'} A(x,x') - 2\xi B(x,x') ;
\end{equation}
whereas equations (3.21) and (3.29) of the same reference read, respectively, as
{\allowdisplaybreaks\begin{eqnarray} 
{G^{(1)}}^t_{\ t'} &=& - \Delta \Phi G^{\rm flat} - 2 \partial_{tt'} A
+ B, \nonumber \\
{G^{(1)}}^t_{\ a'} &=& ( \partial_{t'a} - \partial_{ta'} ) A, 
\nonumber \\ 
{G^{(1)}}^a_{\ t'} &=& \bigl( \partial^a_{\ t'} - \partial^{a'}_{\ t}
\bigr) A,   
\nonumber \\
{G^{(1)}}^a_{\ b'} &=& \delta^a_{\ b} \bigl( \Delta \Phi G^{\rm flat}
- 2 \partial_{tt'} A  - B \bigr) + 
\bigl( \partial^{a'}_{\ b} - \partial^a_{\ b'} \bigr) A , \nonumber \\
\langle \Phi \rangle &\equiv& \Phi(x) + \Phi(x')  
\end{eqnarray} }
and
{\allowdisplaybreaks\begin{eqnarray}
{G^{(1)}}^{tt}_{\ \ t't'} &=& - (\Delta \Phi G^{\rm flat} 
+ \partial_{tt'} A), \nonumber \\
{G^{(1)}}^{tt}_{\ \ t'a'} &=& (\partial_{t'a} - \partial_{ta'}) A, 
\nonumber \\
{G^{(1)}}^{tt}_{\ \ a'b'} &=& -\delta_{ab} \bigl[ 
\langle \Phi \rangle G^{\rm flat} 
+ \partial_{tt'} A + 2B \bigr] 
+ (\partial_a + \partial_{a'}) (\partial_b + \partial_{b'}) A, 
\nonumber \\
{G^{(1)}}^{ta}_{\ \ t't'} &=& \bigl(\partial^a_{\ t'} -
    \partial^{a'}_{\ t} \bigr) A, \nonumber \\
{G^{(1)}}^{ta}_{\ \ t'b'} &=& -\delta^a_{\ b}( \partial_{tt'} A  
  + B) + \frac{1}{2} \bigl(\partial^a_{\ b} 
  + 2\partial^{a'}_{\ b} + \partial^{a'}_{\ b'} \bigr) A, \\
{G^{(1)}}^{ta}_{\ \ b'c'} &=& \delta^a_{\ (b} \bigl(\partial_{c)t'} 
  - \partial_{c')t} \bigr) A, \nonumber \\
{G^{(1)}}^{ab}_{\ \ t't'} &=& \delta^{ab} \bigl[ 
\langle \Phi \rangle G^{\rm flat} - \partial_{tt'} A \bigr] 
+ \bigl(\partial^a + \partial^{a'} \bigr)
  \bigl(\partial_b + \partial_{b'} \bigr) A, 
\nonumber \\
{G^{(1)}}^{ab}_{\ \ t'c'} &=& \delta^{(a}_{\ c} 
\bigl(\partial^{b)}_{\ t'} - \partial^{b')}_{\ t} \bigr) A, 
\nonumber \\  
{G^{(1)}}^{ab}_{\ \ c'd'} &=& \bigl( 2\delta^{(a}_{\ c} 
   \delta^{b)}_{\ d} - \delta^{ab} \delta_{cd} \bigr) 
   \bigl(\Delta \Phi G^{\rm flat} - \partial_{tt'} A \bigr) 
+ \delta^{ab}(\partial_c + \partial_{c'})(\partial_d 
+ \partial_{d'}) A  
\nonumber \\ & & \mbox{} 
- 2\delta^{(a}_{\ (c} \Bigl( \partial^{b)}_{\ d)} 
+ 2 \partial^{b)}_{\ d')} + \partial^{b')}_{\ d')} \Bigr) A 
+ \delta_{cd} \bigl(\partial^a + \partial^{a'} \bigr)
  \bigl(\partial^b + \partial^{b'}\bigr)A 
- 2 \delta^{ab} \delta_{cd} B. \nonumber \\
\Delta \Phi &\equiv& \Phi(x) - \Phi(x') .
\end{eqnarray}}
As noted in \cite{Pfenning:2000zf}, the $G^\text{flat}$ portion of $G^{(1)}$ will not contribute to the self-force, because it is non-zero only on the null cone.

We review the steps of solving the equations (\ref{scalarf}), (\ref{EMf}) and (\ref{gravf}) by introducing the common building blocks of the scalar, electromagnetic and gravitational self force, the 2-point functions $A$ and $B$, 
\begin{equation} \label{AG}
A(x,x')= \frac{1}{2 \pi} \int G^\text{flat}(x,x'')  \Phi (\vec{x}'') G^\text{flat} (x'',x') \dd^4 x''.
\end{equation}
Equation (\ref{AG}) is nothing more than the first Born approximation. The particle emits a signal at the point $x'$, this in turn interacts with the gravitational potential at the point $x''$ and then returns back to the particle at the new location $x$.

Similarly for the density distribution
\begin{equation}\label{BG}
B(x,x')= \frac{1}{2 \pi} \int G^\text{flat}(x,x'')  \rho (\vec{x}'') G^\text{flat} (x'',x') \dd^4 x'',
\end{equation}
Inside the integrals in (\ref{AG}) and (\ref{BG}), 
the $G^\text{flat}(x,x'')$ picks out the past light cone of $x$, while the  $G^\text{flat}(x'',x')$ picks out the future light cone of $x'$. Substituting (\ref{delta}) into (\ref{AG}) and (\ref{BG}), and denoting $\Delta t \equiv t-t'$,
\begin{equation}\label{A}
A(x,x')= \frac{1}{2 \pi} \int \frac{ \Phi (\vec{x}'') }{|\vec{x}-\vec{x}''| |\vec{x}''-\vec{x}'| }  \delta(\Delta t - |\vec{x}-\vec{x}''|- |\vec{x}''-\vec{x}'|) \dd^3 x'',
\end{equation}
and
\begin{equation}\label{B}
B(x,x')= \ \frac{1}{2 \pi} \int \frac{ \rho (\vec{x}'') }{|\vec{x}-\vec{x}''| |\vec{x}''-\vec{x}'| }  \delta(\Delta t - |\vec{x}-\vec{x}''|- |\vec{x}''-\vec{x}'|) \dd^3 x''.
\end{equation}
The delta function in (\ref{A}) and (\ref{B}) enforces the relation,
\begin{equation}
|\vec{x}-\vec{x}''| + |\vec{x}''-\vec{x}'| = \Delta t.
\end{equation}
This defines a two-dimensional surface formed by the intersection of the past light cone of $x$ and the future light cone of $x'$. 
The locus of this surface in $3-$space is an ellipsoid of revolution centered at 
\begin{equation}
\vec{x}_0=\frac{1}{2} (\vec{x}+\vec{x}'),
\end{equation}
of semi-major axis
\begin{equation}
s=\frac{\Delta t}{2},
\end{equation}
and half the inter-focal-distance
\begin{equation}
e=\frac{1}{2} |\vec{x}-\vec{x}'|=\frac{1}{2} R
\end{equation}
We represent the vector $\vec{x}''$ 
as the sum 
of a vector pointing from the origin to the center of the ellipsoid 
and a vector pointing from the center of the ellipsoid to a point on its surface, $\vec{\eta}$,
\begin{equation}
\label{xpp}
\vec{x}''=\vec{x}_0 + \vec{\eta}(s,\theta,\phi)\,.
\end{equation}
We define $\eta_0$ to be a vector from the center of the ellipsoid to the center of the mass distribution.
For convenience, 
we choose the origin at the center of the mass distribution,
$\vec{x}_0 =-\vec{\eta}_0 $. 
Now,
\begin{equation} \label{changecoor}
\vec{x}''= \vec{\eta}-\vec{\eta}_0
\end{equation}
points from the center of the mass distribution
to the surface of the ellipsoid. 

The parametric equations of $\vec{\eta}$ are given by
\begin{align}\label{eta}
\eta_{1} &=\sqrt{s^2 -e^2} \cos \phi \sin \theta, & \eta_{2}&=\sqrt{s^2 -e^2} \sin \phi \sin \theta, & \eta_{3}&=s \cos \theta.
\end{align}
Similarly,
\begin{align}\label{eta0}
\eta_{01}&=\sqrt{s_0^2 -e^2} \cos \phi_0 \sin \theta_0 & \eta_{02}&=\sqrt{s_0^2 -e^2} \sin \phi_0 \sin \theta_0 & \eta_{03} &=s_0 \cos \theta_0 \,,
\end{align}
where
\begin{equation}
s_0=\frac{1}{2} (r+r')
\end{equation}
for $r=\vert {\vec x}\vert$ and $r'=\vert {\vec x'}\vert$.

For the special case of
\begin{align}
\Phi(\vec{x}'') =- \frac{M}{|\vec{x}''|},
\end{align}
we substitute into (\ref{A}) to obtain,
\begin{align}\label{Adetail}
A(x,x')=-\frac{M}{4 \pi} \int \frac{1}{|\vec{\eta}-\vec{\eta}_0|} d\Omega.
\end{align}
The final results are,
\begin{align}\label{Atail}
A(x,x')=-\frac{M}{R}\Theta (\Delta t -R) 
  &\bigg(\Theta (r+r'-\Delta t) \log\frac{r+r'+R}{r+r'-R} \nonumber\\
  &+ \Theta (\Delta t - r - r') \log\frac{\Delta t+R}{\Delta t-R} \bigg),
\end{align}
and 
\begin{equation}
B(x,x')=\frac{M}{r r'} \delta(\Delta t - r - r').
\end{equation}
These expressions have been  derived previously in \cite{Pfenning:2000zf}. 
In examining (\ref{Atail}), 
we notice that it contains two parts. 
The first one, corresponds to the $\Delta t <r+r'$ 
(i.e., $s<s_0$) case. 
This is the early piece of the tail and  
represents the part of the ellipsoid 
before intersecting the mass distribution. 
The late-time tail, which represents the case where 
the ellipsoid has swept past the mass distribution, 
is given by the second piece, $\Delta t > r+r'$
(i.e., $s>s_0$). 
Here we notice an abrupt change in the behaviour of the function, due to the point-like nature of the mass distribution.
For later convenience, 
we give the expressions for $A_\text{early-pt}$ and $A_\text{late-pt}$ 
in terms of the variables $e$, $s$ and $s_0$,
\begin{align} \label{lateandearly}
A_\text{early-pt}=- \frac{M}{2 e} \log\frac{s_0 + e }{s_0 - e}, && A_\text{late-pt}=- \frac{M}{2 e} \log\frac{s + e }{s - e}.
\end{align}
Thus
\begin{equation}
\label{Apt}
A_\text{pt} = 
\begin{cases}
 A_\text{early-pt}, & \text{for  $s<s_0$};\\
 A_\text{late-pt}, & \text{for  $s>s_0$}\,.
 \end{cases}
\end{equation}
The point like nature of the mass distribution is the reason that $A_{pt}$ is continuous but not differentiable across $s=s_0$. As we shall see below, the inclusion of finite size effects will smooth out this transition somewhat. This is the motivation for the following analysis, since in order to properly capture the finite-size effects, we must smooth out the central singularity.

\section{On the interior of the mass distribution} \label{sectionthird}

We start our analysis by considering the Newtonian potential, which is a solution to Poisson's equation
\begin{align}
\label{NewtonianPhi}
\Phi(\vec{x}) 			&= \int_{\mathbb{R}^3} \dd^3 \vec{x}' \frac{\rho(\vec{x}')}{4\pi |\vec{x}-\vec{x}'|} , \\
-\vec{\nabla}^2 \Phi 	&= \rho .
\end{align}
The Green's function of the Laplacian $(4\pi |\vec{x}-\vec{x}'|)^{-1}$ may be expanded as follows,
\begin{align}
\frac{1}{4\pi |\vec{x}-\vec{x}'|} 
&= \frac{1}{r_>} \sum_{\ell=0}^\infty \sum_{m=-\ell}^{\ell} \frac{Y_\ell^m(\widehat{x}) \overline{Y_\ell^m}(\widehat{x}')}{2\ell+1} \left( \frac{r_<}{r_>} \right)^\ell .
\end{align}
Here $r_>$ is the larger of the $(r \equiv |\vec{x}|, r' \equiv |\vec{x}'|)$; and $\widehat{x} \equiv \vec{x}/r$, $\widehat{x}' \equiv \vec{x}'/r'$. This formula implies eq. \eqref{NewtonianPhi} can be written as
\begin{align}
\label{MultipoleExpansion}
\Phi(\vec{x}) &= \int_{0}^{+\infty} \dd r' r'^2 \int_{\mathbb{S}^2} \dd\Omega_{\widehat{x}'} 
\frac{\rho(r',\widehat{x}')}{r_>} \sum_{\ell=0}^\infty \sum_{m=-\ell}^{\ell} \frac{\overline{Y_\ell^m}(\widehat{x}) Y_\ell^m(\widehat{x}')}{2\ell+1} \left( \frac{r_<}{r_>} \right)^\ell .
\end{align}
If $\vec{x}$ lies well outside the matter source, i.e., $\rho(\vec{x}) = 0$, then $r_> = r$ and we have
\begin{align}
\label{MultipoleExpansion_ObserverOutside_I}
\Phi(\vec{x}) &= \sum_{\ell=0}^\infty \sum_{m=-\ell}^{\ell} \frac{Y_\ell^m(\widehat{x})}{r^{1+\ell} (2\ell+1)} 
\int_{0}^{+\infty} \dd r' r'^{2+\ell} \int_{\mathbb{S}^2} \dd\Omega_{\widehat{x}'} \rho(r',\widehat{x}') \overline{Y_\ell^m}(\widehat{x}') .
\end{align}
In this case where $\vec{x}$ lies well outside the matter source, we may also Taylor expand the Green's function
\begin{align}
\label{TaylorExpansion}
\frac{1}{4\pi |\vec{x}-\vec{x}'|} 
= \frac{1}{4\pi r} + \sum_{\ell=1}^\infty \frac{(-)^\ell x'^{i_1} \dots x'^{i_\ell}}{\ell !} \partial_{i_1} \dots \partial_{i_\ell} \frac{1}{4\pi r} \,.
\end{align}
Comparing (\ref{MultipoleExpansion_ObserverOutside_I}) and (\ref{TaylorExpansion}), we see that
\begin{align}
\label{MultipoleExpansion_ObserverOutside_II}
\Phi[\vec{x}] 	&= \frac{M}{4\pi r} + \sum_{\ell=0}^{\infty} \frac{(-)^\ell}{\ell !} M^{i_1 \dots i_\ell} \partial_{i_1} \dots \partial_{i_\ell} \frac{1}{4\pi r}, \\
M 				&\equiv \int_{\mathbb{R}^3} \dd^3\vec{x}' \rho(\vec{x}'), \\
M^{i_1 \dots i_\ell} &\equiv \int \dd^3\vec{x}' \rho(\vec{x}') x'^{i_1} \dots x'^{i_\ell} .
\end{align}
Equations \eqref{MultipoleExpansion_ObserverOutside_I} and \eqref{MultipoleExpansion_ObserverOutside_II} are equivalent.

However, when $\vec{x}$ lies {\it inside} the matter source these formulas are no longer valid. To see their breakdown, simply put $\vec{x}=\vec{0}$ and notice how equations \eqref{MultipoleExpansion_ObserverOutside_I} and \eqref{MultipoleExpansion_ObserverOutside_II} blow up at $r=0$; whereas the actual Newtonian potential inside, say, a uniform spherical mass distribution is most definitely not singular at the origin. In actuality, the $r_>$ and $r_<$ in the Green's function formula tells us, when $\vec{x}$ lies inside the matter source, the integration over $r'$ in (\ref{MultipoleExpansion}) needs to be split in two 
\begin{align}
\label{MultipoleExpansion_ObserverInside}
\Phi(\vec{x}) 
&= \int_{r}^{+\infty} \dd r' r'^2 \int_{\mathbb{S}^2} \dd\Omega_{\widehat{x}'}
\frac{\rho(r',\widehat{x}')}{r'} \sum_{\ell=0}^\infty \sum_{m=-\ell}^{\ell} \frac{\overline{Y_\ell^m}(\widehat{x}) Y_\ell^m(\widehat{x}')}{2\ell+1} \left( \frac{r}{r'} \right)^\ell  \nonumber \\ 
&\qquad
+ \int_{0}^{r} \dd r' r'^2 \int_{\mathbb{S}^2} \dd\Omega_{\widehat{x}'} 
\frac{\rho(r',\widehat{x}')}{r} \sum_{\ell=0}^\infty \sum_{m=-\ell}^{\ell} \frac{\overline{Y_\ell^m}(\widehat{x}) Y_\ell^m(\widehat{x}')}{2\ell+1} \left( \frac{r'}{r} \right)^\ell .
\end{align}
This situation needs to be accounted for in the tail integrals in (\ref{A}) and (\ref{B}).

\subsection{Small-angle approximation}

To capture the finite size effects, and to make progress analytically,
we choose a simple form for the density distribution,

\begin{equation} \label{rho}
 \rho (r)=
    \begin{cases}
   \rho_0 \left(1-\frac{r^2}{\alpha^2} \right)^n& {\mathrm {for}}~r<\alpha, \\
   0 & {\mathrm {for}}~r>\alpha\,,
    \end{cases}
\end{equation}
and fix $n=2$ for simplicity.
This form for $\rho(r)$ was chosen such that it is smooth at both the edges and the center. Furthermore, the motivation for a spherically symmetric (i.e., purely radial) profile can appeal to Birkhoff's theorem. If there were no self-force, the orbiter would experience a spacetime that is sensitive only to the central body's mass. But, we shall shortly see below that the self-force is sensitive to the interior structure. We are now able to perform the integrals in (\ref{MultipoleExpansion_ObserverInside}), to obtain
\begin{equation} \label{phiinandout}
  \Phi(\vec{x}'')=
    \begin{cases}
    - \frac{M}{4\pi\alpha} \frac{105}{48}
    \left(1
          - \frac{|\vec{x}''|^2}{\alpha^2}
          - \frac{1}{7} \frac{|\vec{x}''|^6}{\alpha^6}
          +  \frac{3}{5} \frac{|\vec{x}''|^4}{\alpha^4}
           \right) & {\mathrm {for}}~ |\vec{x}''|<\alpha, \\
    -\frac{M}{4 \pi |\vec{x}''|} & {\mathrm {for}}~|\vec{x}''|>\alpha\,,
    \end{cases}
\end{equation}
where 
\begin{equation}
M \equiv 4\pi\int_0^\alpha \rho(r) r^2 \dd r 
= \frac{32\pi}{105}\rho_0 \alpha^3\,.
\end{equation}
We  want to use this expression for $\Phi$ in
(\ref{A}) and (\ref{B}) to calculate the 2-point functions $A$ and $B$.
To obtain the limits of integration 
we use the parametric equation of a sphere, 
\begin{equation}\label{sphereparam}
|\vec{\eta}-\vec{\eta}_0|^2= \alpha^2.
\end{equation}
This parametric equation defines the boundary $|\vec{x}''| = \alpha$, which marks the jump in the interior vs exterior behavior of the $\Phi$ in eq. (\ref{phiinandout}). While we were unable to exactly evaluate (\ref{A}) for a generic position of the central mass, 
we can do so for locations $\vec{x}$ and $\vec{x}'$ where the small angle approximation applies, i.e.
where the size $\alpha$ of the central body is small compared to its distance from the center of the ellipsoid.
In this case,
\begin{equation}\label{theta}
 \sin \theta =  \sin \theta_0 +  \cos \theta_0 \chi - \frac{1}{2} \chi^2 \sin \theta_0
 +{\cal{O}}\left(\chi^3\right)\,,
 \end{equation}
 where $\chi\equiv\theta-\theta_0$
 and
 \begin{equation}\label{phi}
\cos \phi \approx  1- \frac{\phi^2}{2}
  +{\cal{O}}\left(\phi^4\right)
 \end{equation}
 with $\chi$ and $\phi$ both small angles.
Using  (\ref{eta}), (\ref{eta0}), (\ref{theta}), and (\ref{phi}), we can write 
\begin{equation}\label{etappandstuff}
|\vec{\eta}-\vec{\eta}_0|^2= \eta_{pp}^2 + A_{pp} \chi^2 -B_{pp} \chi + C_{pp} \phi^2 .
\end{equation}
Here, in terms of the variables 
$s$, $s_0$, $\gamma\equiv\sqrt{s^2-e^2}$, $\gamma_0\equiv\sqrt{s_0^2-e^2}$ and $e$, we have defined
\begin{eqnarray}
\eta_{pp}^2 &\equiv&
  (\gamma-\gamma_0)^2 
  + 2 (e^2 -s s_0+\gamma \gamma_0) \cos\theta_0^2,\cr
A_{pp} &\equiv& 
  e^2 + \gamma \gamma_0 
  + (-2 e^2 +s s_0-\gamma \gamma_0) \cos\theta_0^2,\cr
B_{pp} &\equiv& 
  -2 (e^2 -s s_0+\gamma \gamma_0) \cos\theta_0 \sin \theta_0,\cr
C_{pp} &\equiv& \gamma \gamma_0 \sin \theta_0^2.
\end{eqnarray}
Of course, $s$ can be regarded as a function of $\gamma$,
and $s_0$ as a function of $\gamma_0$, 
or {\it vice versa}.

Solving (\ref{sphereparam}), 
the $\phi$ limits of integration are
\begin{align}
\phi_\text{min}&=-\sqrt{\frac{B_{pp} \chi-A_{pp} \chi^2+ \alpha^2-\eta_{pp}^2}{C_{pp}}}, & \phi_\text{max}=+\sqrt{\frac{B_{pp} \chi -A_{pp} \chi^2+ \alpha^2-\eta_{pp}^2}{C_{pp}}},
\end{align}
while the $\chi$ limits of integration are 
\begin{align}
\chi_\text{min}&=\frac{B_{pp}-\sqrt{B_{pp}^2+ 4 A_{pp} (\alpha^2-\eta_{pp}^2)}}{2 A_{pp}}, & \chi_\text{max}=\frac{B_{pp} + \sqrt{B_{pp}^2+ 4 A_{pp} (\alpha^2-\eta_{pp}^2)}}{2 A_{pp}}.
\end{align}
Using (\ref{etappandstuff}) in conjuction with (\ref{A}),  (\ref{changecoor}), (\ref{phiinandout}) and the limits of integration, we calculate the integral over the portion of the surface of the ellipsoid that is interior to the spherical mass distribution, by first integrating with respect to $\phi$  and then with respect to $\chi$,

\begin{eqnarray}
A_\text{interior} 
&=& -M \sin\theta_0
\frac{B_{pp}^2 + 4 A_{pp} (\alpha^2-\eta_{pp}^2)}
{65536 A_{pp}^{9/2} C_{pp}^{1/2} \alpha^7}
\biggl[
      5 B_{pp}^6
      + A_{pp} B_{pp}^4 
          \left(92 \alpha^2 -60 \eta_{pp}^2\right)
 \cr 
  && + 16 A_{pp}^2 B_{pp}^2 
          \left(47 \alpha^4
          - 46 \alpha^2 \eta_{pp}^2
          + 15 \eta_{pp}^4\right)
      \cr
  && + 64  A_{pp}^3
          \left(93 \alpha^6 - 47 \alpha^4 \eta_{pp}^2
          + 23 \alpha^2 \eta_{pp}^4
          -5 \eta_{pp}^6 \right)
    \biggr]    
\end{eqnarray}
We must also evaluate the integral over the portion of the ellipsoid exterior to the mass distribution.
Since we are using the small $\phi$ approximation, we cannot directly compute the exterior piece of $A$. 
However, we observe that the exterior contribution
to the integral is identical to the point-mass case.
Therefore, 
we calculate the interior integral for the point mass
(i.e., with a $1/r$ potential),
\begin{eqnarray}\label{Ainterior-pt}
A_\text{interior-pt}=-M~\frac{ 2 \sqrt{A_{pp}} \alpha + \sqrt{-B_{pp}^2+ 4 A_{pp} \eta_{pp}^2}}{4 A_{pp} \sqrt{C_{pp}}}\,.
\end{eqnarray}
This has two forms: either $A_\text{interior-pt-late}$ or $A_\text{interior-pt-early}$, depending respectively on whether the ellipsoid has or has not swept through the center of the mass distribution. 
We then subtract $A_\text{interior-pt}$ 
from the full-ellipsoid point-mass results  
$A_\text{early-pt}$ or $A_\text{late-pt}$ 
(i.e., (\ref{lateandearly})) as appropriate. 

Following these steps, we write A as a piecewise function,
 \begin{equation}\label{Axx}
 A(x,x')=\begin{cases}
  A_\text{early-pt}, & \text{for  $\chi_\text{range}<0, \gamma<\gamma_0$}.\\
 \left(A_\text{early-pt}-A_\text{interior-pt-early}\right)+A_\text{interior}, & \text{for  $ \chi_\text{range} \geq 0 ,\gamma<\gamma_0$}.\\
  \left(A_\text{late-pt}-A_\text{interior-pt-late}\right)+A_\text{interior}, & \text{for   $\chi_\text{range} \geq 0 , \gamma>\gamma_0$}.\\
 A_\text{late-pt}, & \text{for  $\chi_\text{range}<0 ,\gamma>\gamma_0$}\,,
 \end{cases}
\end{equation}
where 
\begin{eqnarray}
\chi_\text{range} \equiv \chi_\text{max}-\chi_\text{min} =
\left[B_{pp}^2+ 4 A_{pp} (\alpha^2-\eta_{pp}^2)\right]/A_{pp}\,.
\end{eqnarray}
The terms in parentheses represent the exterior contributions to $A$.

\subsection{Simplest case: central body at $\theta_0=\pi/2$}

As an example of our formalism,
we will apply our method first to a simple situation: 
a structureless particle in a circular orbit 
around a spherically symmetric mass distribution.
In this case, $r=r'$, and  
for simplicity we can place the mass distribution (shown in blue in Fig. 1), on the equatorial plane of the ellipsoid given by the coordinate $s_0$, $\theta_0$, and $\phi_0$ (and parametrized by $e$), as in (35). This is the solid ellipse shown in the three panels of Fig. 1, with center at $\theta_0=\pi/2$, $\phi=0$.   The surface of integration, given by the coordinate $s$, $\theta$ and $\phi$ is shown as the dashed ellipses in Fig. 1. At early times (middle panel) the mass distribution lies outside the surface of integration, while at late times (right panel) the mass distribution lies inside it. Equation  (\ref{Axx}) can be rewritten as
\begin{equation}
 A(x,x')=\begin{cases}
 A_\text{early}, & \text{for  $\gamma < \gamma_0-\alpha$}.\\
 A_\text{early-pt}-A_\text{interior-$\frac{\pi}{2}$-pt-early}+A_\text{interior-x}, & \text{for  $\gamma_0 -\alpha < \gamma<\gamma_0$}.\\
 A_\text{late-pt}-A_\text{interior-$\frac{\pi}{2}$-pt-late}+A_\text{interior-x}, & \text{for  $\gamma_0  < \gamma<\gamma_0+\alpha$}.\\
 A_\text{late}, & \text{for $ \gamma>\gamma_0+\alpha$}\,,\\
 \end{cases}
\end{equation}
\begin{figure}[!htb]
\begin{center}
\includegraphics[width=2in]{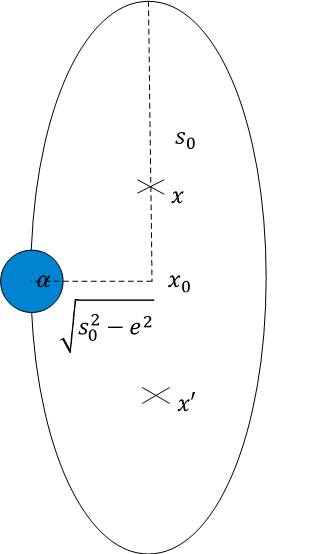}
\includegraphics[width=2in]{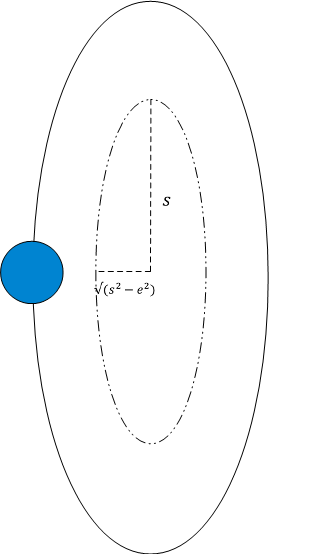}
\includegraphics[width=2in]{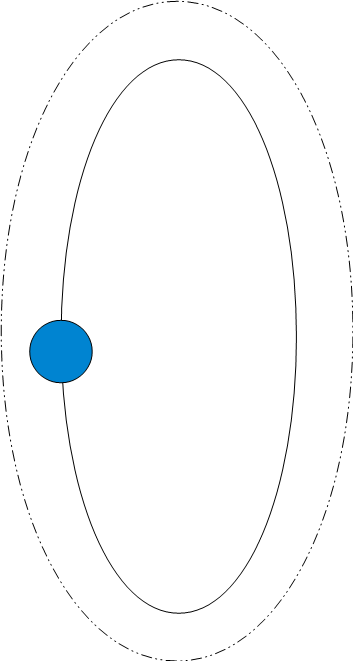}
\end{center}
 \caption[]%
{{\small The leftmost figure denotes the position of the sphere in the ellipsoidal coordinate system used in the paper; see equations \eqref{xpp}, \eqref{eta}, and \eqref{eta0}. The dotted ellipsoid in the middle and rightmost figures represents the ellipsoid of integration in equations \eqref{A} and \eqref{B}. The middle figure describes the early time tail (prior to intersection) while the rightmost one describes the late time tail (after intersection).}}  
\label{fig:1}
\end{figure}
with the various component functions having much simpler forms:
\begin{equation}
A_\text{interior-$\frac{\pi}{2}$-pt-early}
=\frac{-M}{2\sqrt{\gamma \gamma_0 (e^2 + \gamma\gamma_0)}} 
(\alpha + \gamma - \gamma_0)\,, 
\end{equation}
\begin{equation}
A_\text{interior-$\frac{\pi}{2}$-pt-late}=\frac{-M}{2 \sqrt{\gamma \gamma_0 (e^2 + \gamma\gamma_0)}}  (\alpha - \gamma + \gamma_0)\,,
\end{equation}
and
\begin{eqnarray}
A_\text{interior-$\frac{\pi}{2}$}=&&\frac{-M }{
 2  \sqrt{\gamma \gamma_0 (e^2 + \gamma \gamma_0)}} 
 \frac{\alpha}{128} 
 \Bigl[
    93  
  - 140 \left(\frac{\gamma-\gamma_0}{\alpha}\right)^2 \cr
 && \quad\quad
  + 70 \left(\frac{\gamma-\gamma_0}{\alpha}\right)^4 
  - 28 \left(\frac{\gamma-\gamma_0}{\alpha}\right)^6 
  + 5 \left(\frac{\gamma-\gamma_0}{\alpha}\right)^8 \Bigr] \,.
 \end{eqnarray}

 where $\alpha\ll1$. 
   \label{1}
 \begin{figure}[h]
        \centering
        \begin{subfigure}[b]{0.475\textwidth}
            \centering
            \includegraphics[width=\textwidth]{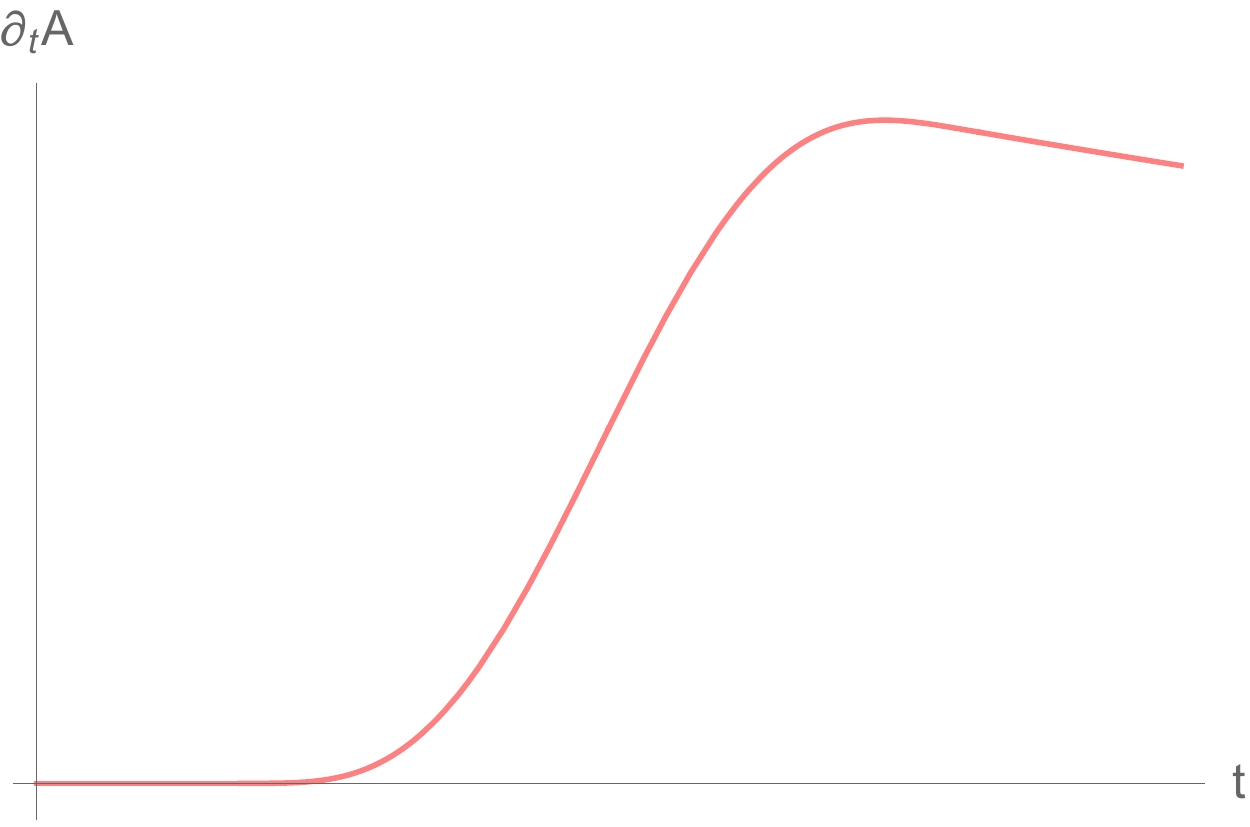}
            \caption[Network2]%
            {{\small $\partial_t  A$ as a function of time}}    
            \label{fig4a}
        \end{subfigure}
        \hfill
        \begin{subfigure}[b]{0.475\textwidth}  
            \centering 
            \includegraphics[width=\textwidth]{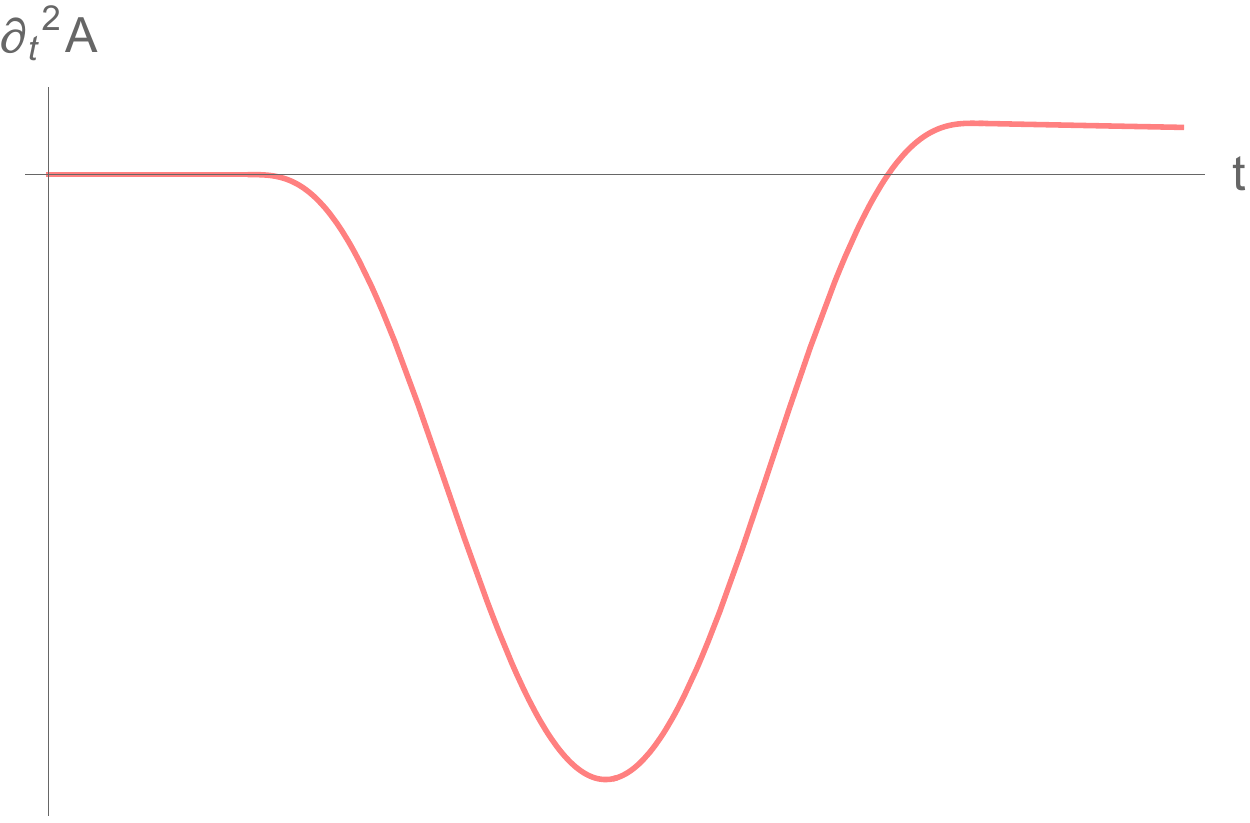}
            \caption[]%
            {{\small $\partial_t ^2 A$ as a function of time}}    
            \label{fig4b}
        \end{subfigure}
        \vskip\baselineskip
        \begin{subfigure}[b]{0.475\textwidth}   
            \centering 
            \includegraphics[width=\textwidth]{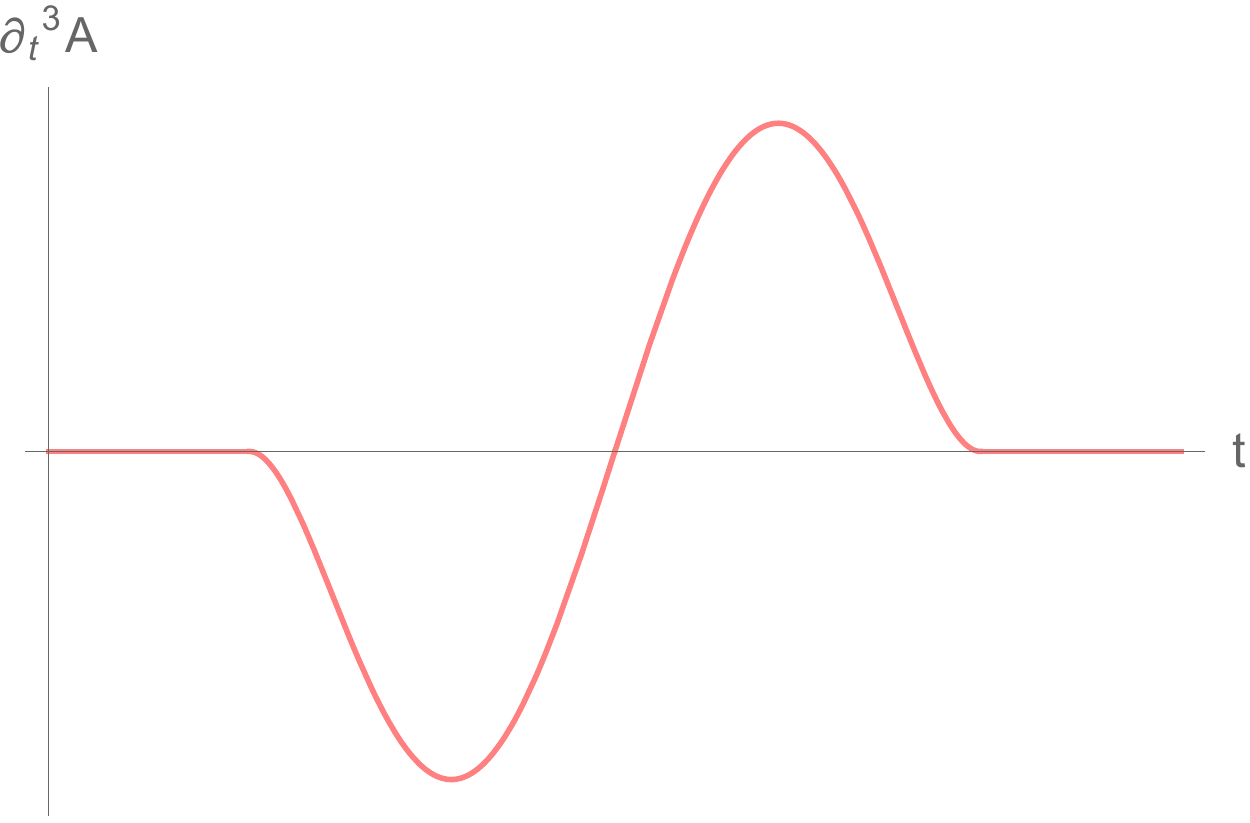}
            \caption[]%
            {{\small $\partial_t ^3 A$ as a function of time}}    
            \label{fig4c}
        \end{subfigure}
        \begin{subfigure}[b]{0.475\textwidth}   
            \centering 
            \includegraphics[width=\textwidth]{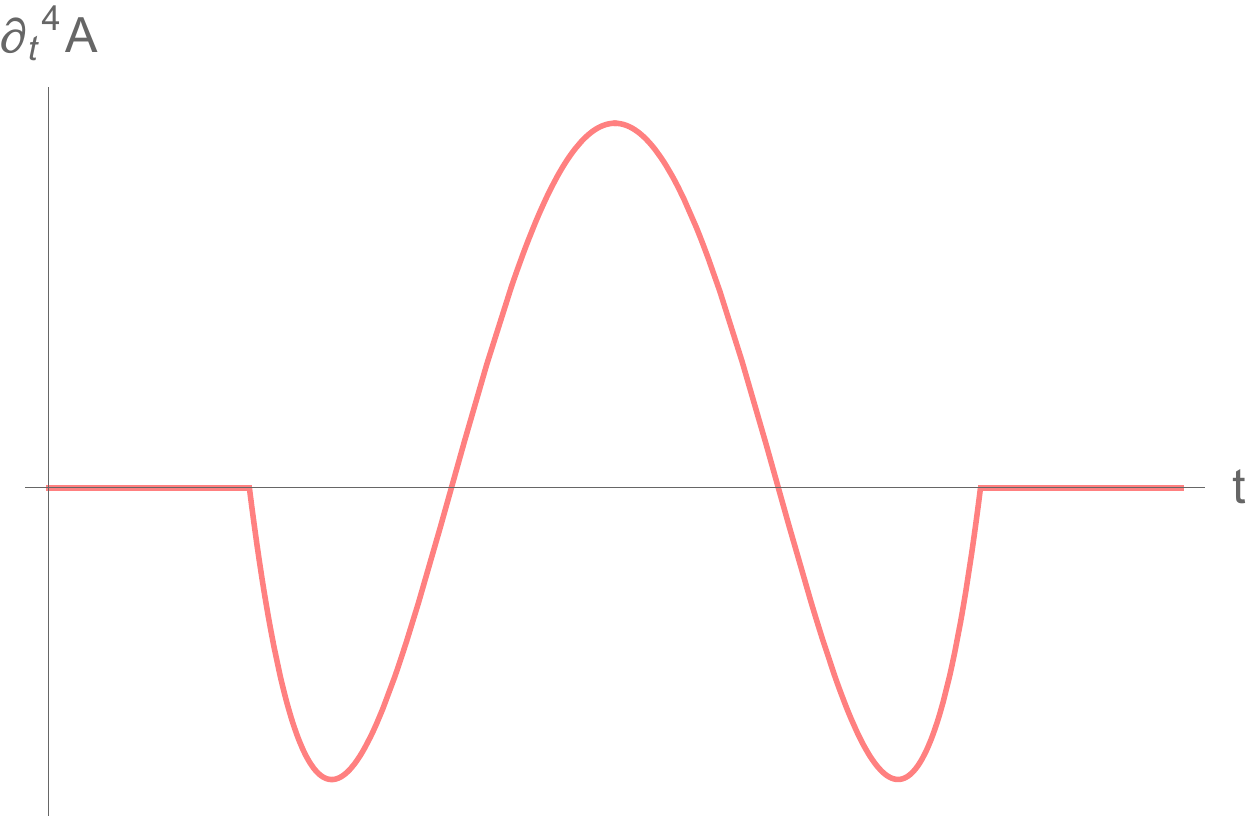}
            \caption[]%
            {{\small $\partial_t ^4 A$ as a function of time}}    
            \label{fig4d}
        \end{subfigure}
        \caption[]
         {  The integrated Newtonian potential as a function of time. We used the values \\ $r =r_p=t_p = 2000$, $R = 1000$, $M = 1$ and $\alpha$=100, to show the qualitative behavior of A and its derivatives. } 
\label{fig:45}
\end{figure}
 
 To obtain a visual picture of the tail, 
 we present the two-point function $A(t)$ and its time derivatives in Fig. \ref{fig:45}. This function appears to be quite sensitive to the smoothness of the density profile of the source of gravity near its surface.
 We chose $n=2$ in (\ref{rho}), so that
 $A$ is three-times-differentiable
 with respect to $t$. This guarantees that we will not encounter any delta function singularities in the self-force.
 
\subsection{B-function}

In a similar way, we construct the B two-point function to first order in the metric perturbation from (\ref{B}),
\begin{equation} 
B=\frac{1}{2} \int { \rho(\vec{x}'') \dd \Omega''}.
\end{equation}
This yields
\begin{equation} \label{Ricci}
B= 
\begin{cases}
\frac{35 M (B_{pp}^2 + 4 A_{pp} (\alpha^2-\eta_{pp}^2))^3}{4096 A^{7/2} \sqrt{C_{pp}} \alpha^7},& 
  \mathrm{inside}~ (\vert\gamma-\gamma_0\vert<\alpha) \cr 
0 & \mathrm{outside}~ (\vert\gamma-\gamma_0\vert>\alpha)\,.
\end{cases}
\end{equation}
For $\theta_0 = \frac{\pi}{2}$, (\ref{Ricci}) using (\ref{rho}) becomes,
\begin{equation} \label{Riccix}
B_{ \theta_0 =\frac{\pi}{2}} = 
\begin{cases}\frac{35 M \left(\alpha ^2-(\gamma-\gamma_0) ^2\right)^3}{{64 \alpha ^7 \sqrt{\gamma  \gamma  _{0}}} \sqrt{\gamma  \gamma  _{0}+e^2}},& 
  \mathrm{inside}~ (\vert\gamma-\gamma_0\vert<\alpha) \cr 
0 & \mathrm{outside}~ (\vert\gamma-\gamma_0\vert>\alpha)\,.
\end{cases}
\end{equation}
 
\section{Finite-size effects} \label{sectionfourth}

We now proceed to evaluate the finite-size effects on the self-force in the non-relativistic and weak field limits.

\subsection{Scalar self-force}
  
We first consider the spatial components of the 
scalar self-force. This includes two pieces, 
\begin{eqnarray} \label{fAandB22}
{f_{B}}^{\textit{scalar}~ a}  &=& -2 \xi q^2 \int_{-\infty}^t \left( B_{,a} + v^a B_{,t} - \frac{1}{2} B_{,a} \vec{v}'^2 + B_{,b} v^b v^a \right) \dd t' , \cr
{f_{A}}^{\textit{scalar}~ a}  &=& -2  q^2 \int_{-\infty}^t \left( A_{,t t' a} + v^a A_{,tt't} + A_{,t t' b} v^b v^a - \frac{1}{2} A_{,t t' a} \vec{v}'^2  \right) \dd t', \quad \vec{v}'^2 \equiv v^{i'} v^{i'}  .
 \end{eqnarray}
Here and throughout the rest of this section, we will keep in $f_B$ and $f_A$ all explicit factors of velocity up to quadratic order. Moreover, these expressions were derived using (\ref{eqn:fscalar}),
assuming that $v^2={\cal O}(\Phi)$,
which holds for bound orbits, and in virialized systems. 
For further details on the derivation of (\ref{fAandB22}) see \cite{Pfenning:2000zf} and references therein. 
To enable us to compute 
$f_{B}^{\textit{scalar}~a}$
and $f_{A}^{\textit{scalar}~a}$ simply, 
we choose a counterclockwise circular orbit in the $1-3$ plane,
\begin{align}\label{param1}
x&= b \cos{\frac{v}{b} t}, & z&= b \sin{\frac{v}{b} t}, & x^\prime&= b \cos{\frac{v}{b} t^\prime}, & z^\prime&= b \sin{\frac{v}{b} t^\prime}.
\end{align}

We use (\ref{Riccix}) in combination with (\ref{param1}), 
to derive the radial component of ${f_{B}^{a}} ^{\textit{scalar}}$ 
to the leading first post-Newtonian (1 PN) order,
\begin{equation}\label{fBxx1}
f_{B\parallel}^{\textit{scalar}}=
  \frac{35 \xi}{16} \frac{ q^2 M b^3}{\alpha^6} 
  \left[5  
    -\frac{22}{3} \frac{\alpha^2}{b^2}  
    - \left(1-\frac{\alpha^2}{b^2}\right)^2  
      \left(5  + \frac{\alpha^2}{b^2}\right) 
      \frac{b}{\alpha}\text{arctanh}\left(\frac{\alpha}{b}\right) 
      +  \frac{\alpha^4}{b^4} \right].
\end{equation} 
We computed this worldline integral and carefully derived the angular limits of integration by solving \footnote{We evaluate the self-force at t=0; and therefore in that limit $\gamma-\gamma_0$ is the same as $|\vec{\eta}-\vec{\eta}_0|$.} $\gamma-\gamma_0= \alpha$  for $t'$ and using the appropriate expressions  for $\gamma,\gamma_0$ in terms of $t$ and $R$.  Here, the integrals are evaluated from $t'=- \infty$ to $t'=0$ with the non-zero part of the B function being evaluated between $t'=- 2 (b+\alpha)+ \frac{{\alpha} (b + {\alpha}) }{b} {v}^2+  {\cal{O}}\left(v^4\right)$ and $t'=- 2 (b-\alpha)+ \frac{{\alpha} (-b + {\alpha}) }{b } {v}^2 +  {\cal{O}}\left(v^4\right)$. Notice that the integrand should be expanded accordingly in powers of $\frac{v}{c}$ so that the appropriate terms are kept in every PN order. 
We need to expand (\ref{fBxx1}) 
in powers of $\zeta\equiv \frac{\alpha}{b}$, 
and keep terms only up to ${\cal{O}}(\zeta^2)$ corrections to the leading term,
so that we are consistent with the small-angle approximation we employed in Sec. \ref{sectionsecond} \footnote{$\zeta$ is an independent small parameter.  On the one hand we are working in the weak field approximation, requiring $\alpha\gg GM \simeq (v/c)^2$, where $v$ is the orbital velocity of the perturber around the source ; on the other hand we require $\alpha\ll b$, the semi-major axis of that orbit.}.
As expected, the leading term is the $\alpha$-independent one, and
\begin{equation}\label{fBxscalar}
f_{B\parallel}^{\textit{scalar}}=2 \xi  q^2\frac{M}{b^3} \left(1+ \frac{2}{9} \zeta^2+ {\cal{O}}\left(\zeta^4\right)\,\right) \,.
\end{equation}
There are, of course, 
${\cal{O}}\left(\xi q^2 M v^2/b^3\right)$
corrections to this, but they are order 2 PN.

An alternative way of deriving the coefficients is given in Appendix B for a generic radial-density profile. We note that the multipole expansion in \cite{Pfenning:2000zf} omitted certain contributions to the static self-force which could have been included.  We present an improved version in Appendix B, which indicates that in fact the finite-size corrections do contribute to the calculation of self-forces, even though they are suppressed by factors of the radius of the mass over the radius of the orbit. That the next order term of the static self-force is sensitive to the interior structure, was in fact pointed out in \cite{Drivas:2010wz} and later in \cite{Isoyama:2012in}. Drivas and Gralla\footnote{Their work was based on \cite{Burko:2000yx}.} pointed out finite size effects for the static part of the self-force, nevertheless our results are general for any density profile as shown in appendix B. Moreover, their mode sum representation of the relevant Green's functions do not provide as much insight into where the finite size effects are coming from, as far as the causal structure of the signal is concerned. Here, we present a more insightful derivation in terms of the tail integral. In addition, we are able to prove concretely that not only the static part of the self-force is sensitive to finite size corrections, but the non conservative part is sensitive as well, meaning that the radiation emitted will carry information about the internal structure of the central body. The radial component of the self-force will not contribute to the power radiated, nevertheless these corrections would affect the orbital evolution of the system.

Similarly, to leading order in $v$,
the tangential-component of $f_{B}^{\alpha \textit{scalar}}$ is
\begin{align}
f_{B\perp}^{\textit{scalar}}= 
  \frac{35 \xi}{16} \frac{q^2 M b^3}{\alpha ^6} 
  {v} 
& \bigg[
     1
    -\frac{8}{3} \zeta^2
    +\frac{11}{5} \zeta^4 
    -\frac{16}{35} \zeta^6
    - \frac{1}{2\zeta} \left(1-\zeta^2\right)^3 
      \log\left(\frac{1+\zeta}{1-\zeta}\right)
  \bigg]\,,
 \end{align}
 
with the perturbative result, 
\begin{equation}
  f_{B\perp}^{\textit{scalar}}=
  2 \xi q^2\frac{M}{b^3} \left( \frac{1}{18} \zeta^2+  {\cal{O}}\left(\zeta^4\right)\right) v\,.
\end{equation}
This contributes to order 1.5 PN. To the same order, the tangential-component of $f_A^{\textit{scalar}}$ contributes as well, with 
\begin{equation}\label{fay}
 f_{A\perp}^{\textit{scalar}}=- q^2\frac{ M}{3 b^3} v \,.
\end{equation}
The integral for the A part of the self-force is evaluated from $t'=-\infty$ to $t'=- 2 (b+\alpha)+ \frac{{\alpha} (b + {\alpha}) }{b} {v}^2$ for late times, from the latter to $t'=- 2 b$ for the middle late, from $t'=- 2 b$ to $t'=- 2 (b-\alpha)+ \frac{{\alpha} (-b + {\alpha}) }{b } {v}^2 $ for the middle early and from the latter to 0 for the early piece. 
Looking at (\ref{fay}), it is quite interesting that the finite-size corrections (which carry the information about the choice of the mass distribution) cancelled perfectly for the part of the self-force that is related to the potential. 
This is consistent mathematically with the result obtained in \cite{Pfenning:2000zf}, though there it was argued that at this PN order, the evaluation of any integrals is unnecessary since the integral can be massaged to a boundary term. 
Here, we have shown this explicitily by directly computing the integral, showing that $f_A^{scalar}$, the part of the self-force that is related to the integrated Newtonian potential, does not receive finite-size corrections at leading order. Finite-size effects do enter into $f_B^{scalar}$. This persists for the electromagnetic case, as we show below. The total scalar self-force at 1.5 PN, ignoring the terms to ${\cal{O}} \left( \zeta^4 \right)$ is given by,
\begin{equation}\label{faying}
 f_{tot\perp}^{\textit{scalar}}=q^2\frac{ M}{ 3 b^3}\left(-1 +  \frac{1}{3}  \xi \zeta^2 \right) v \,.
\end{equation}
 
In the limit of $\zeta \rightarrow 0$,
we recover at 1.5PN the result  obtained in the literature,
\begin{equation}
  {\vec f}^{scalar}=
    2  \xi q^2 \frac{M}{r^3} \hat{r} 
    + \frac{1}{3} q^2 \frac{d {\vec{g}}}{dt}.
\end{equation}

For completeness, we report that the 2PN contributions
to the scalar self-force are not incurred in $f_A$, but do enter $f_B$ -- including finite size corrections:\footnote{Note that we have not included corrections that are of order $M^2$.}
\begin{equation}\label{fBx2PNscalar}
  f_{B\parallel}^{\textit{scalar}}=
- \xi q^2 \frac{M}{b^3} v^2 
  \left(1+ \frac{1}{3} \zeta^2 + {\cal{O}}\left(\zeta^4\right)\right)\,.
\end{equation}
  
 \subsection{Electromagnetic self-force}
 
Just like the scalar, we find 
that the electromagnetic  self-force (\ref{eqn:fem})
has $A$ and $B$ contributions:
 \begin{eqnarray} \label{fA}
{f_{A}^{{\textit{EM}}~a}} =&& e^2\int_{-\infty}^t 
\Bigl[
\left(A_{,t t a '} +  A_{,tt'a}\right) 
+  \left(A_{,t  a ' b} 
-  A_{,t a b'}\right) v^b  
\nonumber\\
&&
+  \left(2 \delta_{a b} A_{,t  t ' t} 
-  A_{,t a ' b} 
+  2 A_{,t a b '} 
-  A_{,t' a b }\right) v^{b '}
\nonumber\\
&&
-2 A_{,t t' a} v^b v_{b'}
+ A_{,c b' a} v^{c'} v^{b}
- A_{a' b c} v^c v^{b'} 
+ 2 A_{t t' b} v^b v^{a'}
\nonumber\\
&&
+ \frac{1}{2}A_{t t' a} v^2
+ \frac{1}{2}A_{t t a'} v^2
\Bigr] \dd t'.
\end{eqnarray}
and
  \begin{equation}\label{fB}
 {f_{B}^{{\textit{EM}}~a}} = 
  -e^2  \int_{-\infty}^t 
  \left(B_{,a } - v^{a'} B_{,t} - B_{,b} v^b v^{a'} + B_{,a} v_{b }v^{b'} + \frac{1}{2} B_{,a} \vec{v}^2   \right) \dd t' .
 \end{equation}
We use (\ref{Riccix}) in combination with (\ref{param1}) to derive the radial-component of $f_{B}^{\textit{EM}}$ to 1 PN order,
\begin{equation}\label{fBxelec}
f_{B\parallel}^{\textit{EM}}=e^2 \frac{M}{b^3} \left(1+ \frac{2}{9} \zeta^2+{\cal{O}}\left(\zeta^4\right)  \right)\,.
\end{equation}
Again, there are ${\cal{O}}\left(e^2 M v^2/b^3\right)$ corrections to this, but they are order 2 PN. Similarly, we calculate the tangential-component (i.e. parallel to the orbital velocity) of $f_{B}^{\textit{EM}}$, 
 \begin{equation}\label{fByelec}
f_{B\perp}^{\textit{EM}}=e^2 \frac{M}{b^3} \left( \frac{1}{18} \zeta^2+  {\cal{O}}\left(\zeta^4\right)\right) v \,,
 \end{equation}
 which contributes to order 1.5 PN. 
 To the same order, 
 \begin{equation}\label{fAyelectr}
f_{A\perp}^{\textit{EM}}=-\frac{2 M}{3 b^3} e^2 v\,.
 \end{equation}
 The total electromagnetic self-force at 1.5 PN, ignoring the terms to ${\cal{O}} \left( \zeta^4 \right)$, is given by,

\begin{equation}\label{fayingem}
 f_{tot\perp}^{\textit{EM}}=e^2\frac{ 2M}{ 3 b^3}\left(-1 +  \frac{1}{12}  \zeta^2 \right) v \,.
\end{equation}

The 2PN contributions to the electromagnetic self-force
do include finite-size effects as well.
For completeness, we report the finite-size corrections to order 2 PN\footnote{Note again that we have not included terms of order $M^2$.} 
for $f_{B}^{\textit{EM}}$ and $f_{A}^{\textit{EM}}$,
\begin{equation}\label{fBx2PNelec}
  f_{B\parallel}^{\textit{EM}}=
  e^2\frac{M}{b^3} v^2 
  \left(\frac{1}{2}+ \frac{5}{18} \zeta^2 + {\cal{O}}\left(\zeta^4\right)\right)\, ,
\end{equation}
 and
\begin{equation}\label{fAx2PNelec}
  f_{A\parallel}^{\textit{EM}}=
e^2 \frac{M}{b^3} v^2 
      \left(1- \frac{2}{9} \zeta^2 +  {\cal{O}}\left(\zeta^4 \right)\right)\, .
\end{equation}
The total electromagnetic self-force at 2 PN to order M, ignoring the terms to ${\cal{O}} \left( \zeta^4 \right)$ is given by,

\begin{equation}\label{fAx2PNelec1}
  f_{tot\parallel}^{\textit{EM}}=
e^2 \frac{3M}{2b^3}  
      \left(1+\frac{1}{27} \zeta^2 \right)v^2\, .
\end{equation}
These expressions for $f_B^{\textit{EM}}$ and $f_A^{\textit{EM}}$ 
 were derived by directly computing (\ref{fB}) and (\ref{fA}), 
 unlike in \cite{Pfenning:2000zf} 
 where the result was derived 
 using the near-coincidence limit. 
 We notice that in the limit of $\alpha \rightarrow 0$ we recover the result obtained in the literature,
\begin{equation}\label{fem}
 \vec{f}^{\,\textit{EM}}=  e^2 \frac{M}{r^3} \hat{r} +\frac{2}{3} e^2 \frac{d \vec{g}}{dt}.
\end{equation}
We immediately notice that we recover the Abraham-Lorentz force along with the static self-force contribution. The latter was obtained in \cite{Smith:1980aa} and was interpreted as a repulsive force required to hold a charge at rest in the presence of a matter distribution. In eqs. \eqref{fBxelec} and \eqref{fByelec}, we showed how this term would receive finite-size corrections. 

It is worth pointing out that, as we would have expected, $\vert\vec{f}^{\,\textit{EM}}\cdot\vec{v}\vert$ gives the usual Larmor formula\footnote{In the pure electromagnetic case, the Larmor formula shows the classical instability of the hydrogen atom.} 
for the radiated power
\begin{equation}\label{fem}
  P= \frac{ 2 e^2 }{3} a^2\,,
\end{equation}
where $a=v^2/b$ is the centripetal acceleration in the circular orbit. Here, we observe the analogue 
of the usual electromagnetic Larmor radiation for a charged particle moving in the gravitational field 
of a central mass, along with its finite size corrections.

 \subsection{Gravitational self-force}
 Similar to the electromagnetic and scalar case, 
 we can calculate the two parts of the gravitational self-force (\ref{eqn:fgrav}),
\begin{eqnarray}\label{fagravit}
  {f_{A}^{\textit{grav}~a}}=
    &&-2 m^2 \int_{-\infty}^t 
      \Bigl[
        (2 A_{,t t a '} -  A_{, t t' a})
        \nonumber\\
    && + \left( -3 \delta_{a b} A_{,t  t' t} 
          + 2  A_{,t a b}  
          + 4  A_{,t a' b} 
          + 2  A_{,t a' b'}\right) v^b 
      \nonumber\\
    && 
      +  \left(4 \delta_{a b} A_{,t t't}
          - 4  A_{,t a'b} 
          - 2 A_{,t a' b'} 
          + 2 A_{,t a b' }\right) ~v^{b'}  
      \nonumber\\
    && 
     + \frac{1}{2}\vec{v}'^2 A_{,tt'a} + \vec{v}'^2 A_{,tta'} + 2 A_{,a'bc}
     v^b v^c \nonumber\\
     &&
     -4 A_{,a'bc} v^b v^{c'} + 2 A_{,a'bc'} v^b v^c - 2 A_{,a'bc'} v^b
     v^{c'} \nonumber\\
     &&
     -A_{,ab'c'} v^b v^c + 2 A_{,ab'c'} v^b v^{c'} - A_{,ab'c'} v^{b'}
     v^{c'} \nonumber\\
     &&
     + A_{,abc} v^b v^c - A_{,abc} v^{b'} v^{c'} + 4 A_{,abc'} v^{b'} v^c \nonumber\\
     &&
     -2A_{,abc'} v^{b'} v^{c'} - 2 A_{,tt'a} v_{d'} v^d - 2 A_{,tta'} v_{d'} v^d \nonumber\\
     &&
     +2 A_{,tta'} \vec{v}^2 - 5 A_{,tt'b} v^a v^b + 2 A_{,tt'b} v^a v^{b'} \nonumber\\
     &&
     +6A_{,tt'b} v^{a'} v^b - 2 A_{,tt'b} v^{a'} v^{b'} + 2 A_{,ttb'} v^a v^b \nonumber\\
     &&
     -2 A_{,ttb'} v^a v^{b'} - 2 A_{,ttb'} v^{a'} v^b + 2 A_{,ttb'} v^{a'} v^{b'} \Bigl]
    \dd t'\,
 \end{eqnarray}
and
\begin{equation} \label{fbgravit}
{f_{B}^{\textit{grav}~a}} 
= -8 m^2  \int_{-\infty}^t \left( v^{a '} B_{,t} + \frac{1}{2} B_{,a} (\vec{v}'^2-2\vec{v}\cdot\vec{v}') + B_{,b} v^b v^{a'} \right) \dd t'.
\end{equation}
 Using the equations for the circular orbit, 
 we obtain from  (\ref{fagravit})
 to order 1.5 PN, 
 for the tangential component of the force
\begin{equation}\label{fAygrav}
  f_{A\perp}^{\textit{grav}}
    = \frac{11 M}{3 b^3}  m^2 v \,.
\end{equation}
Eq. (\ref{fAygrav}) has been previously derived 
  in \cite{Pfenning:2000zf}.
As they observed,  
$f_{A\perp}^{\textit{grav}}$ has the opposite sign 
as its  order 1.5 PN electromagnetic counterpart 
$f_{A\perp}^{\textit{EM}}$ (obtained in (\ref{fAyelectr})), 
 suggesting problematic radiation antidamping. 
 
 The resolution of this puzzle is tied to the fact that, the MiSaTaQuWa self force eq. (\ref{waveeq.grav}) was derived by assuming the background spacetime is completely devoid of matter. To account for the presence of matter, such as our central mass, Pfenning and Poisson \cite{Pfenning:2000zf} demonstrated the need to introduce a 'matter mediated' force $f_{mm}^{grav}$. To leading order, this 'matter mediated' force would in fact cancel the above antidamping self-force in eq. (\ref{fAygrav}),
\begin{equation}\label{fmm}
  f_{mm\perp}^{\textit{grav}}
    = -\frac{11 M}{3 b^3}  m^2 v \,.
\end{equation}
$f_{mm\perp}^{\textit{grav}}$ 
was not introduced {\it {ad hoc}}.
It arises from the fact 
that the finite-mass central body is not fixed, 
so one should simultaneously 
solve the equations of motion 
for the orbiting particle and the central mass distribution. 

There are no 1.5PN corrections to $f_B$ and $f_A$. Whereas at 2PN order, there are no finite size effects in the radial component of $f_B$,
 \begin{equation}\label{fBx2PN}
f_{B\parallel}^{\textit{grav}}=4 m^2\frac{M}{b^3} v^2 \bigg(1- {\cal{O}}\left(\zeta^4\right)\bigg) \, ;
\end{equation}
but there are non-trivial ones occurring in the radial component of (\ref{fagravit}),
 \begin{equation}\label{fAx2PN}
f_{A\parallel}^{\textit{grav}}=4 m^2\frac{M}{b^3} v^2 \left(2- \zeta^2 - {\cal{O}}\left(\zeta^4\right)\right) .
\end{equation}
These are the order $M v^2$ 2PN corrections. 
There will be additional order $M^2$ 2PN corrections,
but these are higher order in our perturbation theory,
and reserved for a future work.

Although for the gravitational case
there are no contributions to the radiated power 
at order 1.5 PN, 
we were able to provide a clear method 
for systematically deriving the finite-size corrections
to the gravitational self-force
by directly calculating the relevant integrals. 
In (\ref{fBx2PN}) and (\ref{fAx2PN}), 
we showed the corrections that enter 
the conservative part of the self-force 
to order 2PN at order $M$. 
These finite-size corrections to the radiated power 
may also enter at 2.5 PN, making it unclear whether or how they would be cancelled
through the matter mediated force.
This will be addressed in a future work.

\section{Summary} \label{Conclusions}
In this work, 
we computed the finite-size corrections 
to the self-force,
in a system involving a structureless particle 
orbiting a finite-size central mass distribution. 
The first corrections enter at 1 PN 
for the scalar and electromagnetic self-force. 
For the gravitational case, our results suggest finite-size effects enter only at order 2PN. However, because the MiSaTaQuWa self-force equation was derived for vacuum spacetimes; as argued by Pfenning and Poisson, we need to follow up our current work with the computation of a 'matter-mediated' force; so as to properly account for the central mass distribution, in order to obtain the complete 2PN gravitational self-force. Nonetheless, we have presented a concrete way of calculating these finite-size corrections to self-forces, by assuming a model for the density distribution and directly computing the appropriate integrals.
Our starting point was the Green's function method 
already used in \cite{Pfenning:2000zf}
and generalized in \cite{Chu:2011ip}. 
We calculated the building blocks, 
namely the potential-two-point function $A$ (in eq. \eqref{AG})
and the density-two-point function $B$ (in eq. \eqref{BG}). This allowed us to see the sensitivity of the A-function
to the smoothness
of the central mass distribution  at its surface. 
By picking a simple mass distribution,
we were able to do all the calculations analytically,
allowing us to demonstrate clearly 
the magnitude and nature of the finite-size effects. 
As a check, 
we compared our findings 
with a previous similar analysis \cite{Pfenning:2000zf}. In the appendix, 
we provide the corrected version of that calculation,
and verify that the methods then match.

In the future, 
we are interested in extending our results 
to higher orders in the Post-Newtonian expansion. 
To do so, 
we need to go to second order 
both in  $M$ and in perturbation theory. 
Although that sounds formidable, 
it is a necessary step 
to fully capture the finite-size effects 
for the gravitational case, 
and to find the first imprints on the radiated power. These finite-size corrections are expected to enter at 2.5 PN order and our aim is to address the possible appearance of anti-damping radiation. From a physical point of view, an application of our results would be the inspiral phase of a neutron star-black hole system \footnote{The black hole would need to have a mass much smaller than the NS, such as might occur if the PBH were a primordial black hole}. In this situation, we might be able to employ the self-force as a probe of the neutron star's internal structure.   However, in truth in this paper we are at least equally interested in the in-principle finite-size effects that we have been able to demonstrate analytically with the specific approximations that we needed to make in order to carry out the calculation. 

\begin{acknowledgments}
KP and GDS would like to thank Craig J. Copi for many useful discussions on the ellipsoidal coordinates used in the paper as well as advice on numerical checks of the results in Fig.(\ref{fig:45}). KP would also like to thank Robert Wald and Leo Stein for suggesting useful references during the Midwest Relativity Meeting. YZC is supported by the Ministry of Science and Technology of the R.O.C. under the grant 106-2112-M-008-024-MY3. KP and GDS are supported by the Department
of Energy grant DE-SC0009946 to the particle astrophysics theory group at CWRU.
\end{acknowledgments}

\appendix
\section{Changing coordinates}

The small-$\phi$ approximation works well for
angles close up to the z-axis.
However, if we want to study what happens 
where that the approximation breaks down, 
we have to change coordinates. 
We use instead the following modified parametric equations:
\begin{align}
\eta_{01}&= s_0 \cos \phi_0 \sin \theta_0, & \eta_{02}&=\sqrt{s_0^2 -e^2} \sin \phi_0 \sin \theta_0, & \eta_{03}=\sqrt{s_0^2 -e^2} \cos \theta_0,
\end{align}
for the center of the ellipsoid, and
\begin{align}
\eta_{1}& =s \cos \phi \sin \theta, & \eta_{2}&=\sqrt{s^2 -e^2} \sin \phi \sin \theta, & \eta_{3}=\sqrt{s^2 -e^2}  \cos \theta,
\end{align}
for the points along the surface. 
Using $(\ref{sphereparam})$, we obtain the new coefficients,
\begin{equation}
\eta_{pp}'^2=(s-s_0)^2 - 2 (e^2 -s s_0+\gamma \gamma_0) \cos\theta_0^2,
\end{equation}

\begin{equation}
A_{pp}'=-e^2 + s s_0 + (- e^2 +s s_0+\gamma \gamma_0) \cos\theta_0^2,
\end{equation}

\begin{equation}
B_{pp}'=-2 (e^2 -s s_0+\gamma \gamma_0) \cos\theta_0 \cos \theta_0,
\end{equation}

\begin{equation}
C_{pp}'=(s s_0-e^2) \sin \theta_0^2.
\end{equation}

\section{An alternative way to calculate 
the static self force}

We wish to evaluate the static limit of eq. (5.19) of \cite{Pfenning:2000zf}:
\begin{align}
\label{fBa}
f_B^i \equiv -e^2 \frac{\partial}{\partial x^i} \int_{-\infty}^{t-0^+} B\left(x=(t,\vec{z}),x'=(t',\vec{z})\right) \dd t';
\end{align}
where $\vec{z}$ is the {\it time-independent} trajectory 
of the point particle; 
and, from eq. (4.2) of \cite{Pfenning:2000zf},
\begin{align}
  \label{B1}
  B(x,x')
  &= \int \dd^3 \vec{x}'' 
    \rho(\vec{x}'') \frac{\delta[t-t'-|\vec{x}-\vec{x}''|-|\vec{x}''-\vec{x}'|]}{|\vec{x}-\vec{x}''||\vec{x}''-\vec{x}'|} .
\end{align}
When both $\vec{x}$ and $\vec{x}'$ 
lie outside the gravitational source $\rho(\vec{x}'')$, 
we may Taylor expand the factor multiplying $\rho$ 
in powers of $\vec{x}''$. 
The zeroth-order term 
would be proportional to the total mass 
$M \equiv \int \rho(\vec{x}'') \dd^3 \vec{x}''$; 
the 1st-order term would be proportional to
the center-of-mass $C^i \equiv \int \rho(\vec{x}'') x''^i \dd^3 \vec{x}''$; the second-order term is proportional to the (time-independent) quadrupole moment
\begin{align}
Q^{ij} \equiv \int x''^i x''^j \rho(\vec{x}'') \dd^3 \vec{x}'' .
\end{align}
Note that all the terms proportional 
to derivatives of $\delta[t-t'-r-r']$ in eq. \eqref{B}, 
arising from the Taylor expansion, 
would yield zero when plugged into eq. \eqref{fB}. 
For, due to the time-independent character 
of the rest of the integrand, 
a typical term would be
\begin{align}
  Q^{ij} \Pi^{ij} \int_{-\infty}^{t-0^+} 
    \partial_{t'}^n \delta[t-t'-r-r'] \dd t' 
  = Q^{ij} \Pi^{ij} \partial_{t'}^{n-1} \delta[-r-r'] 
  = 0 ;
\end{align}
for some integer $n > 0$. We have defined
\begin{align}
r \equiv |\vec{x}| 
\qquad \text{ and } \qquad
r' \equiv |\vec{x}'| .
\end{align}
Therefore, for static self-forces, we merely need to focus on Taylor expanding the denominator of eq. \eqref{B1}. In particular,
\begin{align}
\label{Dxx''}
\frac{1}{|\vec{x}-\vec{x}''|}
&= \frac{1}{r} + \frac{\vec{x}'' \cdot \widehat{r}}{r^2} - \frac{1}{2} \frac{x''^i x''^j}{r^3} \left( \delta^{ij} - 3 \widehat{r}^i \widehat{r}^j \right) + \dots \\
\label{Dx'x''}
\frac{1}{|\vec{x}'-\vec{x}''|}
&= \frac{1}{r'} + \frac{\vec{x}'' \cdot \widehat{r}'}{r'^2} - \frac{1}{2} \frac{x''^i x''^j}{r'^3} \left( \delta^{ij} - 3 \widehat{r}'^i \widehat{r}'^j \right) + \dots ; \\
\widehat{r} &\equiv \vec{x}/|\vec{x}|, \qquad \widehat{r}' \equiv \vec{x}'/|\vec{x}'| .
\end{align}
This in turn implies \footnote{In \cite{Pfenning:2000zf}, Pfenning and Poisson replaced the denominator of eq. \eqref{B1} with $rr'$; i.e., without carrying out the expansions in equations \eqref{Dxx''} and \eqref{Dx'x''}. Because eq. (4.17) in \cite{Pfenning:2000zf} only contained derivatives of $\delta(t-t'-r-r')$, as explained above, their result does not capture the static portion of the self-force.} 
\begin{align}
f_B^i 
&= -e^2 \frac{\partial}{\partial x^i} \int_{-\infty}^{t-0^+} \bigg( \frac{M}{r r'}  + \dots \nonumber \\&+ Q^{ab} \left\{ \frac{1}{r} \cdot \frac{\delta^{ab} - 3 \widehat{r}'^a \widehat{r}'^b}{-2 r'^3} + \frac{\delta^{ab} - 3 \widehat{r}^a \widehat{r}^b}{-2 r^3} \frac{1}{r'} + \frac{\widehat{r}^a \widehat{r}'^b}{r^2 r'^2} \right\} + \dots \bigg) \delta[t-t'-r-r'] \dd t' \nonumber \\
\label{fB_Quadrupole}
&= -e^2  \frac{\partial}{\partial x^i} \bigg( \frac{M}{r r'} + \dots \nonumber \\&+ Q^{ab} \left\{ \frac{\widehat{r}^a \widehat{r}'^b}{r^2 r'^2} - \frac{1}{2} \frac{1}{r} \cdot \frac{\delta^{ab} - 3 \widehat{r}'^a \widehat{r}'^b}{r'^3} - \frac{1}{2} \frac{\delta^{ab} - 3 \widehat{r}^a \widehat{r}^b}{r^3} \frac{1}{r'} \right\} + \dots \bigg)_{\vec{x}=\vec{x}'=\vec{z}} .
\end{align}
{\bf Example} \qquad Consider the mass density in eq. \eqref{rho},
\begin{equation}
\rho(\vec{x}'')
= 
  \begin{cases}
  \rho_0 \left( 1 - \left(\frac{r''}{a}\right)^2 \right)^2,
  & {\mathrm{for}}\quad r''\leq a\,,\\
  0\,,& {\mathrm{for}}\quad r''\geq a\,.
  \end{cases}
\end{equation}
Here $r'' \equiv |\vec{x}''|$, and we have implicitly chosen the center-of-mass to be at the origin $\vec{x}''=\vec{0}$. Then a direct calculation yields
\begin{align}
Q^{ab} &= \delta^{ab} \cdot \frac{M a^2}{9} , \\
M &\equiv 4\pi \int_{0}^{a} \rho(r'') r''^2 \dd r'' .
\end{align}
Because $Q^{ab} \propto \delta^{ab}$ only the first term  in the quadrupole contribution of eq. \eqref{fB_Quadrupole} matters.
\begin{align}
f_B^i 
&= -e^2 \left( -\widehat{r}^i \frac{M}{r^2 r'} + \frac{M a^2}{9} \frac{\partial}{\partial x^i} \left\{ \frac{\vec{x} \cdot \vec{x}'}{r^3 r'^3} \right\}_{\vec{x}=\vec{x}'=\vec{z}} + \dots \right) \\
&= -e^2 \left( -\widehat{r}^i \frac{M}{r^3} + \frac{M a^2}{9} \left\{ \frac{x'^i - 3 (\vec{x} \cdot \vec{x}') \widehat{x}^i/r}{r^3 r'^3} \right\} + \dots \right)_{\vec{x}=\vec{x}'=\vec{z}} \\
&= e^2 \frac{M}{b^3} \widehat{r}^i \left( 1 + \frac{2}{9} \frac{a^2}{b^2} + \dots \right) ,
\end{align}
where $b \equiv |\vec{z}| = |\vec{x}| = |\vec{x}'|$. This matches eq. \eqref{fBxelec}.
\newpage
\bibliographystyle{apa}
\bibliography{selfforcepaper_v4}
\end{document}